\documentclass[prd,superscriptaddress,reprint]{revtex4-1}

\usepackage[usenames,dvipsnames]{xcolor}
\usepackage{amsmath}
\usepackage{subcaption}
\usepackage{graphicx}
\usepackage{physics}
\usepackage[colorlinks=true,linkcolor=NavyBlue]{hyperref}
\begin{document}
\title{Exploring the dark matter inelastic frontier with 79.6 days of PandaX-II data}
\date{\today}
\affiliation{INPAC and School of Physics and Astronomy, Shanghai Jiao Tong University, Shanghai Laboratory for Particle Physics and Cosmology, Shanghai 200240, China}
\author{Xun Chen}
\email[Corresponding author: ]{chenxun@sjtu.edu.cn}
\affiliation{INPAC and School of Physics and Astronomy, Shanghai Jiao Tong University, Shanghai Laboratory for Particle Physics and Cosmology, Shanghai 200240, China}
\author{Abdusalam Abdukerim}
\email[Corresponding author: ]{abdusalam291@qq.com}
\affiliation{School of Physics and Technology, Xinjiang University, \"{U}r\"{u}mqi 830046, China}
\author{Wei Chen}
\affiliation{INPAC and School of Physics and Astronomy, Shanghai Jiao Tong University, Shanghai Laboratory for Particle Physics and Cosmology, Shanghai 200240, China}
\author{Yunhua Chen}
\affiliation{Yalong River Hydropower Development Company, Ltd., 288 Shuanglin Road, Chengdu 610051, China}
\author{Xiangyi Cui}
\affiliation{INPAC and School of Physics and Astronomy, Shanghai Jiao Tong University, Shanghai Laboratory for Particle Physics and Cosmology, Shanghai 200240, China}
\author{Deqing Fang}
\affiliation{Shanghai Institute of Applied Physics, Chinese Academy of Sciences, 201800, Shanghai, China}
\author{Changbo Fu}
\author{Karl Giboni}
\affiliation{INPAC and School of Physics and Astronomy, Shanghai Jiao Tong University, Shanghai Laboratory for Particle Physics and Cosmology, Shanghai 200240, China}
\author{Franco Giuliani}
\affiliation{INPAC and School of Physics and Astronomy, Shanghai Jiao Tong University, Shanghai Laboratory for Particle Physics and Cosmology, Shanghai 200240, China}

\author{Xuyuan Guo}
\affiliation{Yalong River Hydropower Development Company, Ltd., 288 Shuanglin Road, Chengdu 610051, China}

\author{Zhifan Guo}
\affiliation{School of Mechanical Engineering, Shanghai Jiao Tong University, Shanghai 200240, China}

\author{Ke Han}
\affiliation{INPAC and School of Physics and Astronomy, Shanghai Jiao Tong University, Shanghai Laboratory for Particle Physics and Cosmology, Shanghai 200240, China}
\author{Shengming He}
\affiliation{Yalong River Hydropower Development Company, Ltd., 288 Shuanglin Road, Chengdu 610051, China}
\author{Xingtao Huang}
\affiliation{School of Physics and Key Laboratory of Particle Physics and Particle Irradiation (MOE), Shandong University, Jinan 250100, China}

\author{Xiangdong Ji}
\email[Spokesperson: ]{xdji@sjtu.edu.cn}
\affiliation{T.D. Lee Institute, Shanghai, China}
\affiliation{INPAC and School of Physics and Astronomy, Shanghai Jiao Tong University, Shanghai Laboratory for Particle Physics and Cosmology, Shanghai 200240, China}
\author{Yonglin Ju}
\affiliation{School of Mechanical Engineering, Shanghai Jiao Tong University, Shanghai 200240, China}

\author{Shaoli Li}
\author{Heng Lin}
\affiliation{INPAC and School of Physics and Astronomy, Shanghai Jiao Tong University, Shanghai Laboratory for Particle Physics and Cosmology, Shanghai 200240, China}

\author{Huaxuan Liu}
\affiliation{School of Mechanical Engineering, Shanghai Jiao Tong University, Shanghai 200240, China}

\author{Jianglai Liu}
\affiliation{INPAC and School of Physics and Astronomy, Shanghai Jiao Tong University, Shanghai Laboratory for Particle Physics and Cosmology, Shanghai 200240, China}
\affiliation{T.D. Lee Institute, Shanghai, China}

\author{Yugang Ma}
\affiliation{Shanghai Institute of Applied Physics, Chinese Academy of Sciences, 201800, Shanghai, China}
\author{Yajun Mao}
\affiliation{School of Physics, Peking University, Beijing 100871, China}
\author{Jinhua Ning}
\affiliation{Yalong River Hydropower Development Company, Ltd., 288 Shuanglin Road, Chengdu 610051, China}
\author{Xiangxiang Ren}
\affiliation{INPAC and School of Physics and Astronomy, Shanghai Jiao Tong University, Shanghai Laboratory for Particle Physics and Cosmology, Shanghai 200240, China}

\author{Fang Shi}
\affiliation{INPAC and School of Physics and Astronomy, Shanghai Jiao Tong University, Shanghai Laboratory for Particle Physics and Cosmology, Shanghai 200240, China}
\author{Andi Tan}
\affiliation{Department of Physics, University of Maryland, College Park, Maryland 20742, USA}

\author{Cheng Wang}
\affiliation{School of Mechanical Engineering, Shanghai Jiao Tong University, Shanghai 200240, China}

\author{Hongwei Wang}
\affiliation{Shanghai Institute of Applied Physics, Chinese Academy of Sciences, 201800, Shanghai, China}

\author{Meng Wang}
\affiliation{School of Physics and Key Laboratory of Particle Physics and Particle Irradiation (MOE), Shandong University, Jinan 250100, China}

\author{Qiuhong Wang}
\affiliation{Shanghai Institute of Applied Physics, Chinese Academy of Sciences, 201800, Shanghai, China}

\author{Siguang Wang}
\affiliation{School of Physics, Peking University, Beijing 100871, China}

\author{Xiuli Wang}
\affiliation{School of Mechanical Engineering, Shanghai Jiao Tong University, Shanghai 200240, China}

\author{Xuming Wang}
\affiliation{INPAC and School of Physics and Astronomy, Shanghai Jiao Tong University, Shanghai Laboratory for Particle Physics and Cosmology, Shanghai 200240, China}

\author{Qinyu Wu}
\affiliation{INPAC and School of Physics and Astronomy, Shanghai Jiao Tong University, Shanghai Laboratory for Particle Physics and Cosmology, Shanghai 200240, China}

\author{Shiyong Wu}
\affiliation{Yalong River Hydropower Development Company, Ltd., 288 Shuanglin Road, Chengdu 610051, China}

\author{Mengjiao Xiao}
\affiliation{Department of Physics, University of Maryland, College Park, Maryland 20742, USA}
\affiliation{Center of High Energy Physics, Peking University, Beijing 100871, China}

\author{Pengwei Xie}
\affiliation{INPAC and School of Physics and Astronomy, Shanghai Jiao Tong University, Shanghai Laboratory for Particle Physics and Cosmology, Shanghai 200240, China}

\author{Binbin Yan}
\affiliation{School of Physics and Key Laboratory of Particle Physics and Particle Irradiation (MOE), Shandong University, Jinan 250100, China}

\author{Yong Yang}
\affiliation{INPAC and School of Physics and Astronomy, Shanghai Jiao Tong University, Shanghai Laboratory for Particle Physics and Cosmology, Shanghai 200240, China}

\author{Jianfeng Yue}
\affiliation{Yalong River Hydropower Development Company, Ltd., 288 Shuanglin Road, Chengdu 610051, China}

\author{Dan Zhang}
\author{Hongguang Zhang}
\affiliation{INPAC and School of Physics and Astronomy, Shanghai Jiao Tong University, Shanghai Laboratory for Particle Physics and Cosmology, Shanghai 200240, China}

\author{Tao Zhang}
\author{Li Zhao}
\affiliation{INPAC and School of Physics and Astronomy, Shanghai Jiao Tong University, Shanghai Laboratory for Particle Physics and Cosmology, Shanghai 200240, China}

\author{Jifang Zhou}
\affiliation{Yalong River Hydropower Development Company, Ltd., 288 Shuanglin Road, Chengdu 610051, China}
\author{Ning Zhou}
\affiliation{INPAC and School of Physics and Astronomy, Shanghai Jiao Tong University, Shanghai Laboratory for Particle Physics and Cosmology, Shanghai 200240, China}
\author{Xiaopeng Zhou}
\affiliation{School of Physics, Peking University, Beijing 100871, China}

\collaboration{PandaX-II Collaboration}

\begin{abstract}
  We report here the results of searching for inelastic scattering of
  dark matter (initial and final state dark matter particles differ by
  a small mass splitting) with nucleon with the first 79.6-day of
  PandaX-II data (Run 9). We set the upper limits for the
  spin independent WIMP-nucleon scattering cross section up to a mass
  splitting of 300 keV/c$^2$ at two benchmark dark matter masses of 1
  and 10 TeV/c$^2$.
\end{abstract}
\maketitle

After more than 30 years of experimental effort, Weakly Interactive
Massive Particles (WIMPs) remain hidden in most recent dark matter
direct detection experiments (see Ref.~\cite{Undagoitia:2015gya} and
\cite{Liu:2017drf} for a recent review).  One possible explanation for
the null results in these experiments is that the elastic scattering
between the WIMPs and nucleon is heavily suppressed. In some
theoretical
scenarios~\cite{TuckerSmith:2001hy,TuckerSmith:2004jv,Chang:2008gd},
inelastic scattering, in which the dark matter particle converts to an
excited state particle with a small mass splitting between the two,
becomes dominant. This could explain the observed annual modulation in
the DAMA/LIBRA data~\cite{Bernabei:2013xsa}, but the annual modulation
in signal was incompatible with the results from several other
different experiments~\cite{Abe:2015eos,Aprile:2017yea}.  In a recent
study~\cite{Bramante:2016rdh}, the inelastic scenario was reconsidered
in the full range of inelastic mass splitting allowed by
kinematics. The authors argued that the direct detection experiments
employing targets with heavy nuclei could be used to explore the
so-called ``inelastic frontier'' - in which the dark matter mass
splitting is in several hundred keV/c$^2$ range. They hypothesized
that the four events with high nuclear recoil energy identified by the
CRESST experiment~\cite{Angloher:2015ewa} might be due to the WIMP
inelastic scattering.  Following the suggestion in
Ref.~\cite{Bramante:2016rdh}, in this article, we report a search of
the WIMP inelastic scattering using the data from the first physics
run of the PandaX-II experiment from March 9 to June 30, 2016 (Run 9,
79.6 live days).

The PandaX-II detector has been described in detail in
Ref.~\cite{Tan:2016diz}.  The central component is a dual-phase xenon
time-projection-chamber (TPC) with a dodecagonal cross section
confined by polytetrafluoroethylene (PTFE) walls, with a sensitive
xenon target mass of 580 kg. The primary scintillation photons ($S1$)
and the secondary proportional scintillation ($S2$) are detected by
two arrays of Hamamatsu R11410-20 photomultipliers (PMTs) located at
the top and bottom.  In comparison to the elastic WIMP scattering, due
to the mass splitting $\delta$, only WIMPs with sufficiently large
velocity (or kinetic energy) relative to the nucleus could be
scattered inelastically, resulting in a reduced phase space.  A
non-zero minimal recoil energy $E_{\text{R}}^{\text{min}}$ also arises
from the requirement on the minimum kinetic energy.

Suppression of the kinetic phase space in inelastic WIMP-nucleus
scattering can be described by the minimal required relative velocity
$v_{\text{min}}$ for a given recoil energy $E_{\text{R}}$. It is
\begin{equation}
  \label{eq:v_min}
  v_{\text{min}} = \frac{1}{\sqrt{2E_{\text{R}}m_{\text{N}}}}\left( \frac{E_{\text{R}}m_{\text{N}}}{\mu} + \delta \right),
\end{equation}
where $m_{\text{N}}$ is the mass of target nucleus and $\mu$ is the
reduced mass of the system. The global minimal velocity is determined
only by the mass splitting and the reduced mass, 
\begin{equation}
  \label{eq:v_min_eq}
  v_{\text{min}}^{\text{global}} = \sqrt{\frac{2\delta}{\mu}}.
\end{equation}
Therefore, a more massive WIMP will lead to larger $\mu$ and smaller
$v_{\text{min}}^{\text{global}}$, resulting in a lesser suppression of
the phase space.  Since most of the nuclei used for dark matter
detection have a mass on the order of 100 GeV/c$^2$, when WIMP mass is
larger than 10 TeV/c$^2$, $\mu$ approaches $m_{\text{N}}$ and
$v_{\text{min}}^{\text{global}}$ approaches a constant.  Therefore,
we selected 1 TeV/c$^2$ and 10 TeV/c$^2$ as two reference dark matter
masses to perform the analysis, following the detailed discussion in
Ref.~\cite{Bramante:2016rdh}.

\begin{figure}[tb]
  \centering
  \begin{subfigure}[b]{1.0\linewidth}
    \includegraphics[width=0.95\linewidth]{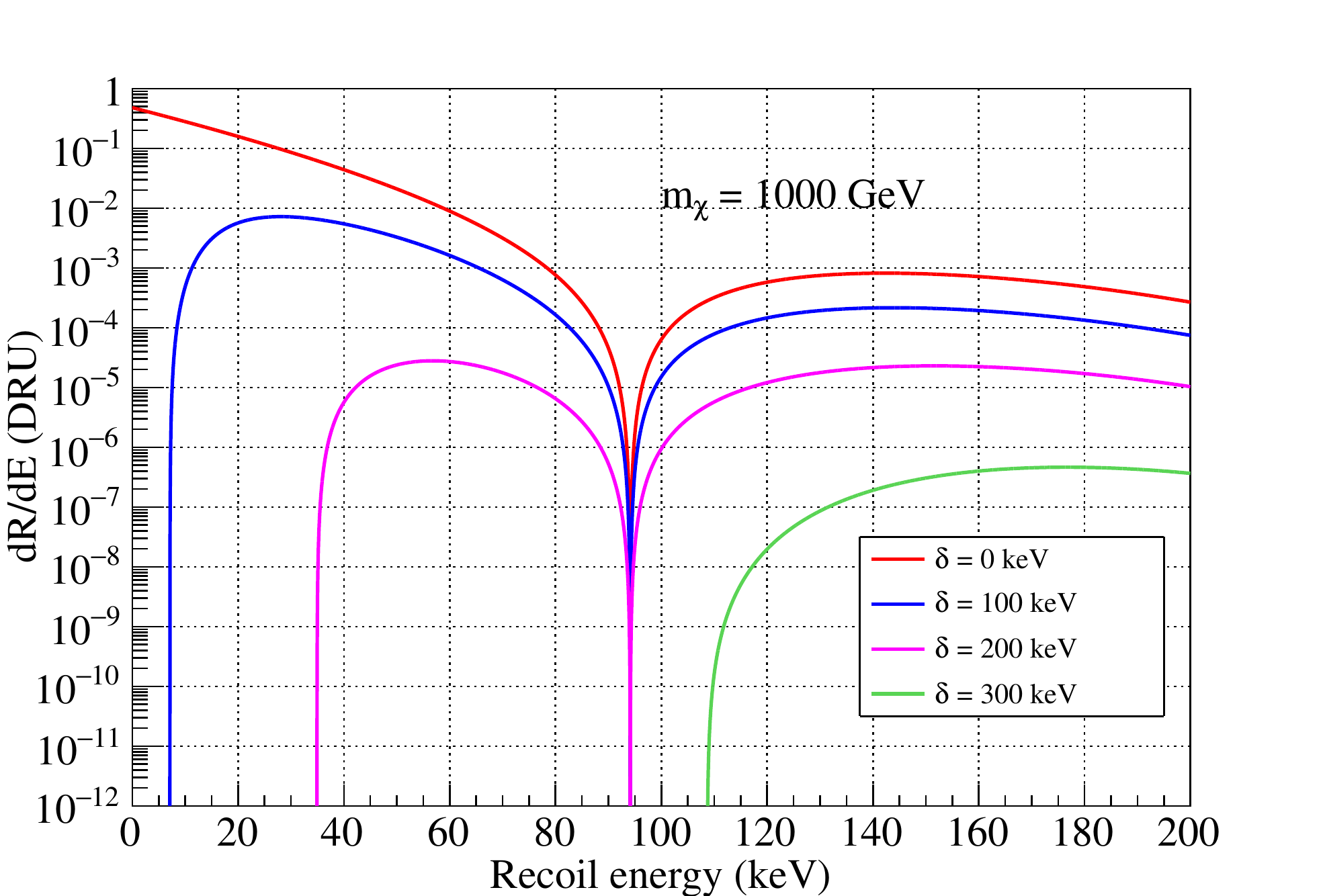}
    \caption{WIMP mass $m_{\chi} = 1$~TeV/c$^2$.}
  \end{subfigure}
  \begin{subfigure}[b]{1.0\linewidth}
    \includegraphics[width=0.95\linewidth]{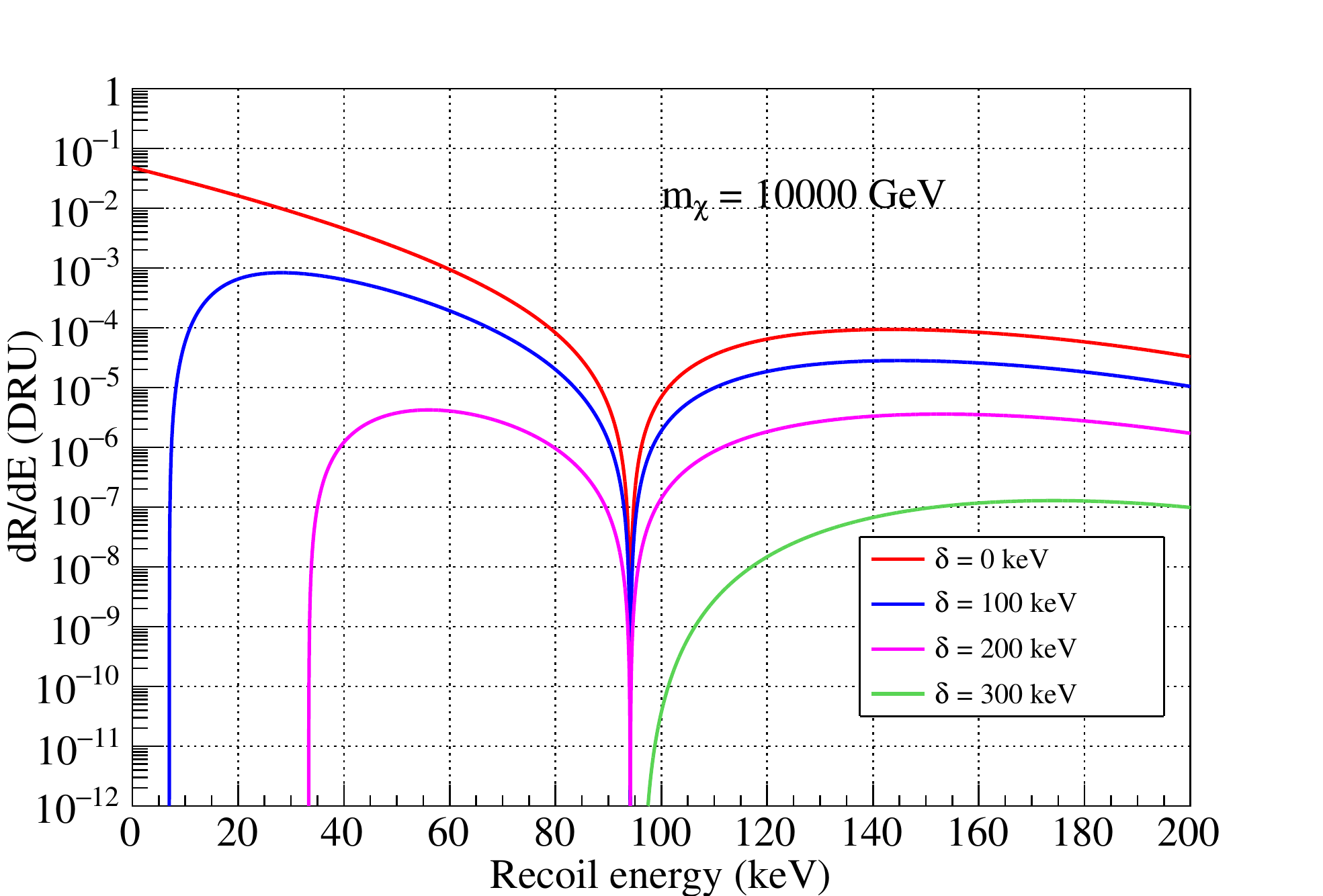}
    \caption{WIMP mass $m_{\chi} = 10$~TeV/c$^2$.}
  \end{subfigure}
  \caption{The expected event rate for scattering between WIMPs and Xe nuclei at the WIMP masses of 1 TeV/c$^2$ (top) and 10 TeV/c$^2$ (bottom), respectively, and the mass splitting values of 0, 100, 200, 300 keV/c$^2$. The SI WIMP-nucleon cross section $\sigma_n$ is fixed at $10^{-40}$ cm$^{2}$. The rate is given in the unit of keV$^{-1}$kg$^{-1}$day$^{-1}$ (DRU).}
  \label{fig:example_event_rate}
\end{figure}

The differential energy spectrum of inelastic scattering is given by
\begin{equation}
  \label{eq:diff_spec_wimp}
  \dv{R}{E} = \frac{\rho_0}{m_{\chi}m_{\text{N}}}\int_{v_{\text{min}}}^{v_{\text{max}}}\dv{\sigma}{E}vf(\vec{v},t)\dd[3]{v},
\end{equation}
where $\rho_0$ is the local density of WIMPs, $m_{\chi}$ is the WIMP
mass, $\dv*{\sigma}{E}$ is the differential scattering cross section,
and $f(\vec{v},t)$ is the WIMP velocity distribution in the earth's
rest frame. The formula has the same form as that of the elastic
scattering, with the only difference coming from the lower limit
$v_{\text{min}}$ of the integral of velocity distribution. The upper
limit $v_{\text{max}}$ is determined by the escape velocity
$v_{\text{esc}}$ of dark matter in the Galaxy. We consider the
DM-nuclei scattering to be spin and energy independent, in which the
differential cross section $\dv*{\sigma}{E}$ takes the form of
\begin{equation}
  \label{eq:si_xsection_detailed}
  \dv{\sigma}{E} = \frac{m_{\text{N}}\sigma_n}{2\mu_{n}^2v^2}\cdot(Z\cdot f^p+(A-Z)\cdot f^n)^2F_{\text{SI}}^2(E).
\end{equation}
The parameter $\mu_n$ is the reduced mass of the WIMP-nucleon system,
and $\sigma_n$ is the spin independent (SI) WIMP-nucleon cross section
at zero momentum transfer in the elastic ($\delta=0$) limit. $f^{n,p}$
are the effective WIMP coupling to neutron and proton,
respectively. $Z$ and $A$ are the atomic and mass number of the target
nucleus. $F_{\text{SI}}^2(E)$ is the SI nuclear form factor, and we
use the Helm parameterization~\cite{Lewin:1995rx} in our
analysis\footnote{Other nuclear structure input,
  e.g. Ref.~\cite{Vietze:2014vsa}, would have little effects for mass
  splitting less than 100 keV/c$^2$, but bring sizable change for masses
  above. For example, the integrated rate for delta=300 keV would
  reduce by a factor of 2.4.}. Three WIMP parameters are required to
calculate the spectrum.  $m_{\chi}$, $\sigma_n$ are explicit in
Eqn.~\ref{eq:si_xsection_detailed}, and the third one, $\delta$,
determines the lower limit $v_{\text{min}}$ of the integral.  With the
standard assumptions $\rho_0=0.3$~GeV/$c^2$/cm$^3$,
$v_{\text{esc}}=544$~km/s, the solar system average circular velocity
$220$~km/s, and the average earth velocity $232$~km/s, we calculate
the expected event rate for inelastic scattering by taking the SI
WIMP-nucleon cross section $\sigma_n$ to be $10^{-40}$~cm$^2$ at the
dark matter mass of 1 TeV/c$^2$ and 10 TeV/c$^2$, respectively, with
four different mass splitting values. The results are shown in
Fig.~\ref{fig:example_event_rate}. The minimal recoil energy is below
40 keV for both dark matter masses when $\delta=200$ keV/c$^2$, so it
could be accessed with the previously published PandaX-II
data~\cite{Tan:2016zwf}.  But for the case that $\delta=300$
keV/c$^2$, the minimal recoil energy is larger than 100 keV/c$^2$ for
$m_\chi=1$ TeV/c$^2$, therefore falling out of the energy range of the
same data selection.

\begin{figure}[tb]
  \centering
  \includegraphics[width=0.95\linewidth]{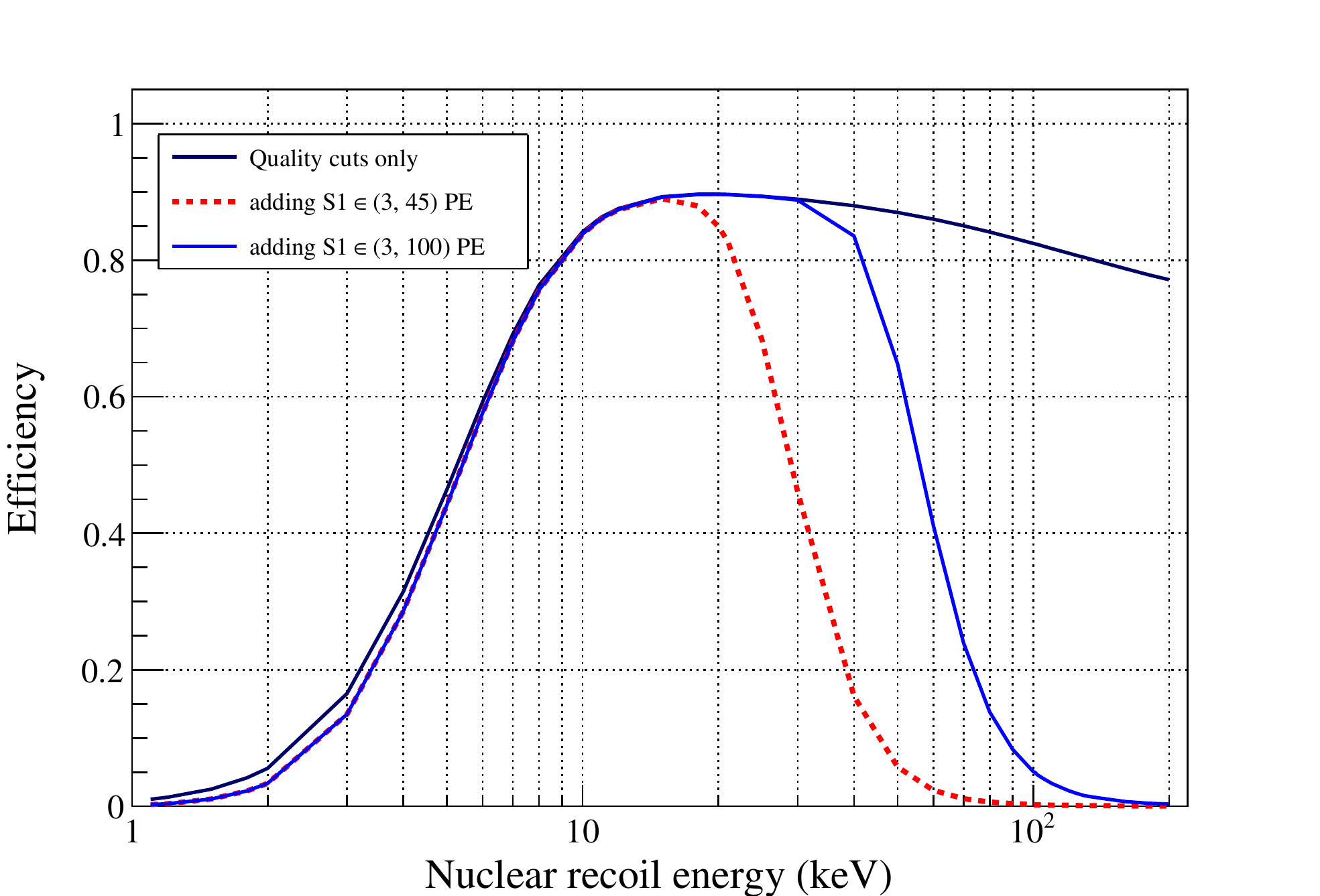}
  \caption{The detection efficiency of PandaX-II experiment as a
    function of the nuclear recoil energy. The dark blue curve
    represents the selection efficency of data quality cuts (see
    Ref.~\cite{Tan:2016zwf}). The red dashed line is obtained with
    additional $S1$ range cut of (3, 45) PE and an upper range of
    10,000 PE for $S2$. The blue solid line is obtained with an
    extended $S1$ and $S2$ window: $S1$ within (3, 100) PE, and S2
    less than 12,000 PE.}
  \label{fig:eff_curve}
\end{figure}

In PandaX-II, the data selection is made by requiring $S1$ and $S2$ be
within a certain range (acceptance). In combination with other data
selection cuts, these correspond to an overall detection efficiency
curve as a function of true nuclear recoil energy.  In the analysis of
SI elastic WIMP-nucleon scattering with PandaX-II
data~\cite{Tan:2016zwf}, the reported average energy window was
between 4.6 to 35.0~keV$_{nr}$ (nuclear recoil energy).  However, due
to the smearing of $S1$ and $S2$, nuclear recoil events with energy up
to 100~keV$_{nr}$ could still enter into the search window, as
indicated in Fig.~\ref{fig:eff_curve}.

\begin{figure}[tb]
  \centering
  \includegraphics[width=0.95\linewidth]{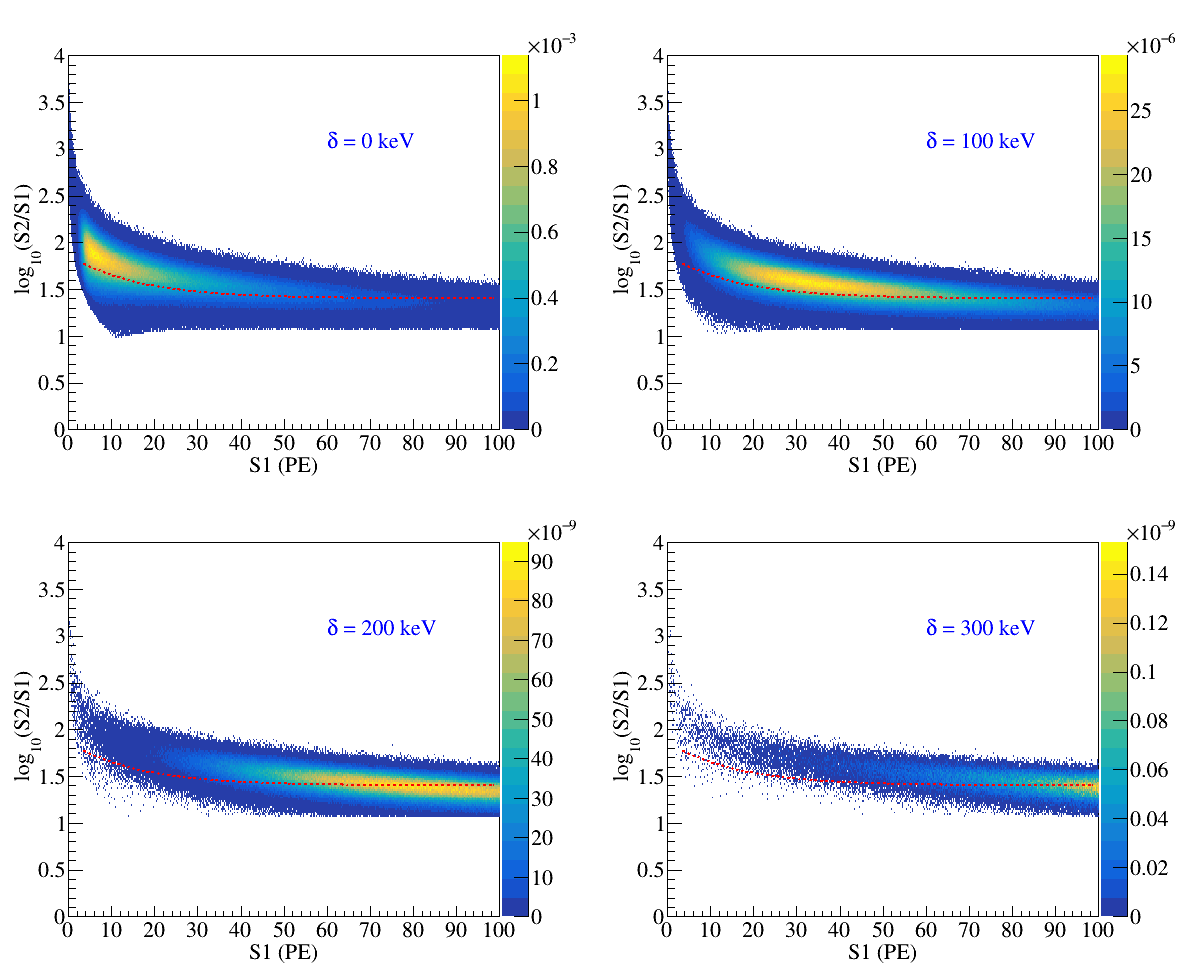}
  \caption{The expected signal distribution of WIMP-nucleon scattering
    at the WIMP mass of $m_\chi=1$ TeV/c$^2$ and the mass splitting
    $\delta$ values of 0, 100, 200 and 300 keV/c$^2$, with one set of
    PandaX-II detector condition parameters. The SI WIMP-nucleon cross
    section $\sigma_n$ is fixed at $10^{-40}$ cm$^{2}$. The total
    expected event rates at each mass splitting are 5.16, 0.21,
    5.23$\times10^{-4}$ and $3.18\times10^{-7}$ evt/kg/day,
    respectively. The $35\%$ quantile of the NR band from AmBe
    calibration (red dashed line) is overlaid in each plot. }
  \label{fig:example_signal_distribution}
\end{figure}

The expected signal distributions of $\log_{10}(S2/S1)$ versus $S1$ for 1 TeV/c$^2$ WIMP with 
different values of $\delta$ in PandaX-II are shown in Fig.~\ref{fig:example_signal_distribution}. Besides the parameters used to calculate the event rate, detector parameters including the photon detection efficiency (PDE), electron extraction efficiency (EEE), single electron gain (SEG), and the electron lifetime, are used to generate the distributions. To simulate
$S1$ and $S2$ given a nuclear recoil energy, the NEST model~\cite{Lenardo:2014cva} was used. At a given value of $\sigma_n$, with the increasing of mass splitting, the expected number of event decreases, and the signal band moves towards higher energy. But a sizable fraction of events can still be found with $S1<$100~PE even when $\delta=300$ keV/c$^2$. 

We followed the identical data selection procedure reported in
Ref.~\cite{Tan:2016zwf}. All data quality cuts remained the same. 
After the same fiducial volume cut, the liquid
xenon mass was estimated to be $329\pm16$~kg. The total exposure in
Run 9 is about $2.6\times10^4$~kg-day.
The final candidate selection was made by expanding the
$S1$ upper window to 100 PE and $S2$ (corrected) upper window to 12,000 PE,
corresponding to an average upper energy of 18.3 keV$_{ee}$ or 68.6 keV$_{nr}$.
Though a higher nuclear recoil energy window has been studied by XENON100~\cite{Aprile:2017aas}, we conservatively choose the value of 100 PE with two considerations. Firstly, the ER background in higher energy region is greatly enhanced by the 33 keV K-shell x ray generated in the $^{127}$Xe electron capture process. Secondly, our ER background model is only optimized in the region with energy up to the end-point decay energy of tritium, i.e., 18.6 keV$_{ee}$.
This improves the detection efficiency for events with
higher nuclear recoil energies, as shown in
Fig.~\ref{fig:eff_curve}. In total, 716 events are observed with the 
expanded signal window. Based on the same background model as in 
Ref.~\cite{Tan:2016zwf}, the expected
total number of background events is 744$\pm$88. The expected background 
composition can be found in Table \ref{tab:bkg_budget}.

\begin{table}[tb]
  \centering
  \begin{tabular}{cccc|c|c}
    \\\hline\hline
 &ER&Accidental&Neutron&\parbox[t]{1.4cm}{Total\\Expected}& \parbox[t]{1.4cm}{Total\\ Observed}\\\hline
    Run 9 & 729.4 & 14.2 & 1.1 & 744$\pm$88 & 716 \\\hline
    \parbox{1.8cm}{Below $35\%$ quantile of\\NR band } & 2.52 & 0.73 & 0.42 &3.7$\pm$0.4 &4\\\hline
  \end{tabular}
  \caption{The expected background events in Run 9 with expanded signal window. The fractional uncertainties of expected events are the same as those in Ref.~\cite{Tan:2016zwf}.}
  \label{tab:bkg_budget}
\end{table}

\begin{figure}[tb]
  \centering
  \includegraphics[width=0.95\linewidth]{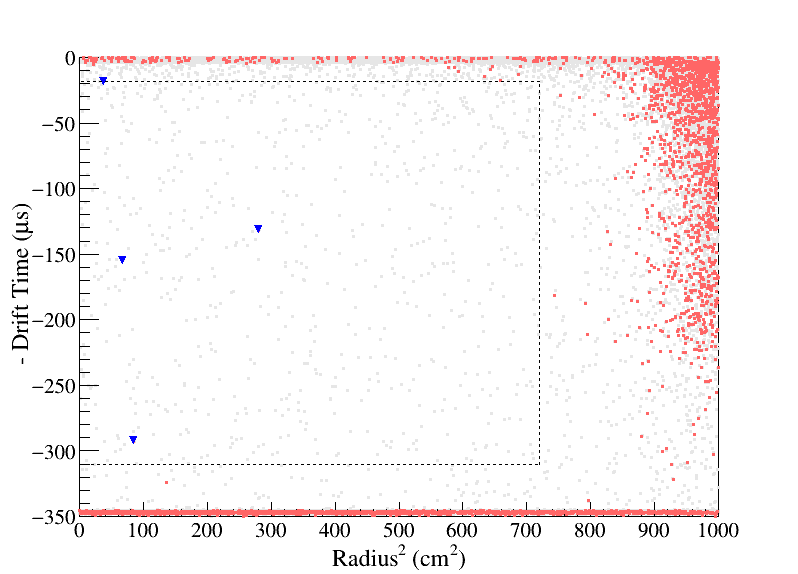}
  \caption{Spatial distribution of events before the FV cut with expanded signal window (gray points), and those of events below the reference acceptance line (outside FV: red points; inside FV: blue triangles). The FV cuts are indicated by the black dashed lines. }
  \label{fig:spatial_dis}
\end{figure}

\begin{figure}[tb]
  \centering
  \includegraphics[width=0.95\linewidth]{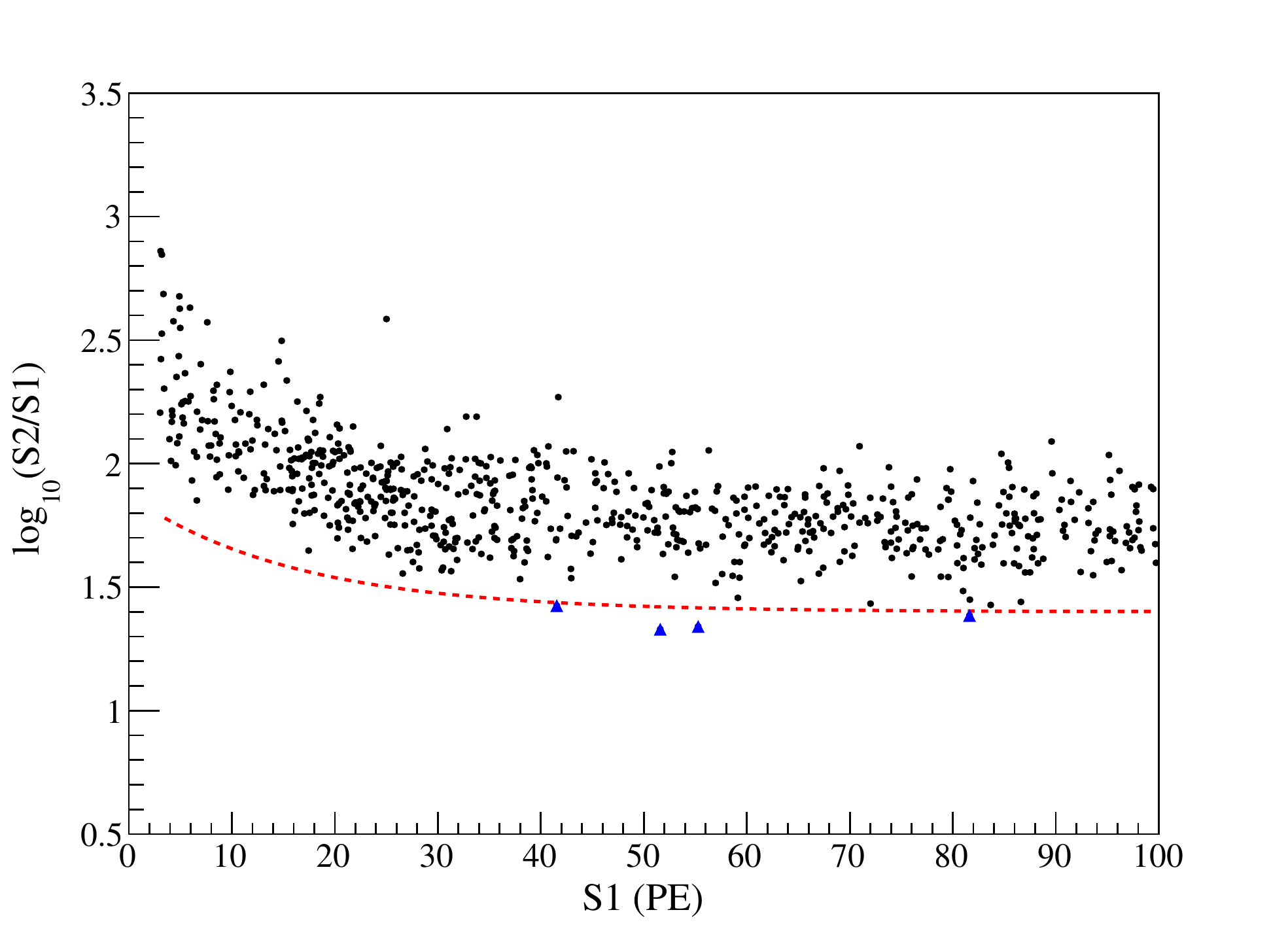}
  \caption{The $\log_{10}(S2/S1)$ vs. $S1$  distribution of selected events within the expanded signal window. A reference acceptance line is indicated as the red dashed line, below 
which four events (blue triangles) are identified.}
  \label{fig:log_dis}
\end{figure}

The spatial distribution of events before and after the FV cut with
the extended signal range is shown in Fig.~\ref{fig:spatial_dis}. The
events show good uniformity within the FV. The $\log_{10}(S2/S1)$
vs. $S1$ distribution of these events is shown in
Fig.~\ref{fig:log_dis}. The reference line is chosen to be the $35\%$ quantile of the calibration AmBe NR band reported in Ref.~\cite{Tan:2016zwf}.
Four events are identified below the line. They are also marked in
Fig.~\ref{fig:spatial_dis}. The event with the smallest $S1$ value was also reported in Ref.~\cite{Tan:2016zwf}. The event with a $S1$ value of 51.8 PE
appeared at the top edge of the FV. After further investigation of its
waveform, we found that this event contains spurious signals which
could be produced by the discharge on the electrodes. The waveforms of the other 
two events are of good quality. The event with a $S1$ value of 
55.0 PE has a larger distance from the reference line, but it is located close to the 
bottom of the FV. 

\begin{figure}[tb]
  \centering
  \begin{subfigure}[b]{1.0\linewidth}
    \includegraphics[width=0.95\linewidth]{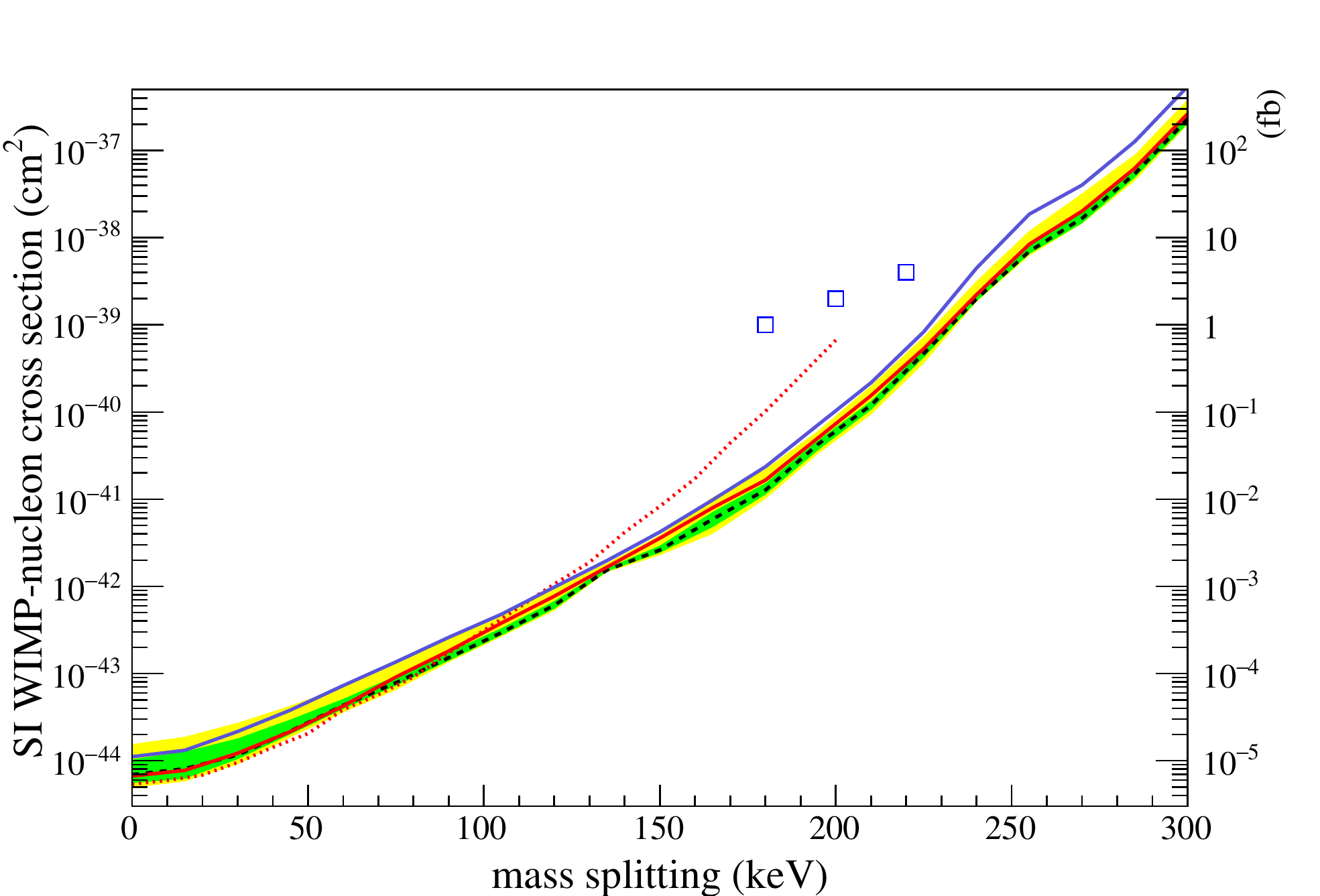}
    \caption{WIMP mass $m_{\chi} = 1$~TeV/c$^2$.}
    \label{limit_1tev}
  \end{subfigure}
  \begin{subfigure}[b]{1.0\linewidth}
    \includegraphics[width=0.95\linewidth]{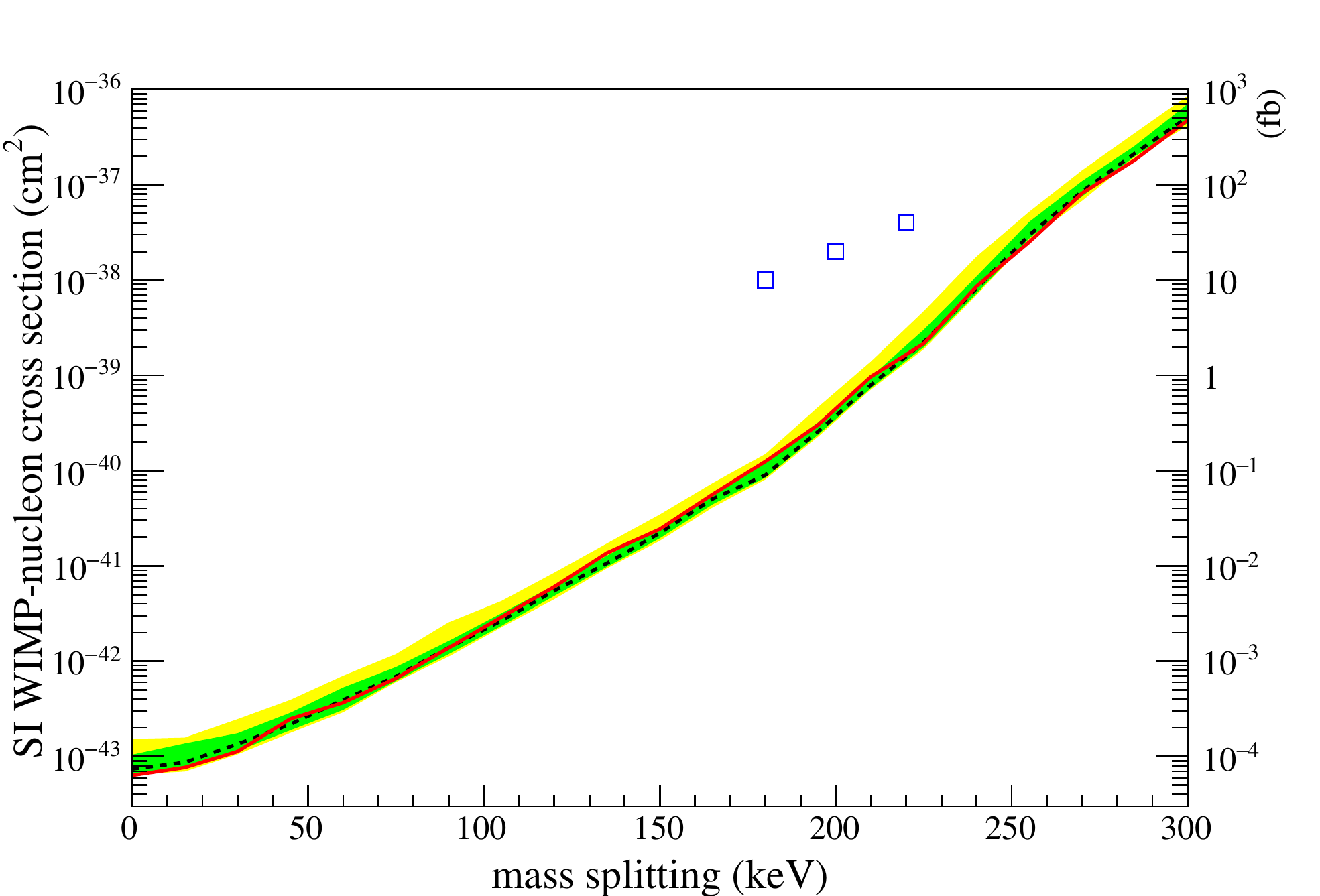}
    \caption{WIMP mass $m_{\chi} = 10$~TeV/c$^2$.}
  \end{subfigure}
  \caption{The 90\% C.L. upper limits (red solid line) for the SI WIMP-nucleon cross section $\sigma_n$ from the PandaX-II run 9 data with expanded signal window at two reference dark matter masses of 1 TeV/c$^2$ (top) and 10 TeV/c$^2$ (bottom). The 1 and 2-$\sigma$ sensitivity bands are shown in green and yellow, respectively, and the medians of the sensitivity band are given in black dashed line. The red dotted line gives the upper limits calculated with the original data selection window. Upper limits calculated with a tuned NEST model by LUX are indicated by the blue solid line. The blue squares indicate the possible WIMP mass splittings and nucleon scattering cross sections by the CRESST data~\cite{Bramante:2016rdh}.}
  \label{fig:fit_results_1tev}
\end{figure}

We used all final 716 candidates to derive the exclusion limits for
the scattering cross section between WIMPs and nucleons at the two
reference dark matter masses of 1 TeV/c$^2$ and 10 TeV/c$^2$. The
profile likelihood fitting was applied to the distribution of $S1$ and
$S2$. We used identical background models from
Ref.~\cite{Tan:2016zwf}. The final $90\%$ confidence level (C.L.)
cross section upper limits were calculated with the CLs
approach~\cite{CLS1, CLS2}, and are shown in
Fig.~\ref{fig:fit_results_1tev}. In both cases, our limit curves lie
between the 1-$\sigma$ and 2-$\sigma$ sensitivity bands when the mass
splitting is within the (100, 200) keV/c$^2$ range, as a result of
three observed events below the reference acceptance line. The limit
calculated with data in the original signal window ($S1$ between 3-45
PE), also shown in the figure, is tighter at lower mass splitting,
since only one event was found below the reference line. To study the
impact of nuclear recoil model to the derived limit, we also plot the
limits calculated based on the ``best-tuned'' NEST from LUX D-D
calibration~\cite{Akerib:2016mzi} for a WIMP-mass of 1 TeV/c$^2$. Due
to reduction of the NR acceptance, the alternative limit is weaker but
not too far beyond the $+1\sigma$ sensitivity band.  In comparison to
the earlier phenomenological treatment (up to about 200 keV/c$^2$ for
LUX-PandaX-II) in Ref.~\cite{Bramante:2016rdh}, this analysis extends
to a wider mass splitting range up to 300 keV/c$^2$. Our results are also
incompatible with the hypothesis that the four high nuclear recoil
events observed by CRESST~\cite{Angloher:2015ewa} were due to
inelastic WIMPs with a mass of 1 TeV/c$^2$ and mass splitting around
200 keV/c$^2$ (which would suggest a cross section at
$\mathcal{O}(10^{-39})$ cm$^2$).

In conclusion, we report the inelastic WIMP search results using
PandaX-II Run 9 data with an exposure of $2.6\times10^4$ kg-day. We
expanded the search signal window and set the $90\%$ upper limits on
the SI WIMP-nucleon cross section for the first time for dark matter
masses of 1 TeV/c$^2$ and 10 TeV/c$^2$ with inelastic mass splittings
up to 300 keV/c$^2$. The search strategy can be further improved with
the continuation of the PandaX-II operation, which will amount to a
factor 4 increase in the total exposure.

\begin{acknowledgments}
  This project has been supported by a 985-III grant from Shanghai
  Jiao Tong University, grants from National Science Foundation of
  China (Nos. 11365022, 11435008, 11455001, 11505112 and 11525522), a
  grant from the Ministry of Science and Technology of China
  (No. 2016YFA0400301). We thank the support of a key laboratory grant
  from the Office of Science and Technology, Shanghai Municipal
  Government (No. 11DZ2260700), and the support from the Key
  Laboratory for Particle Physics, Astrophysics and Cosmology,
  Ministry of Education. This work is supported in part by the Chinese
  Academy of Sciences Center for Excellence in Particle Physics
  (CCEPP).  We would like to thank Dr. Jian-Hui Zhang for the useful
  discussions on the integral of WIMP velocity distribution. We also
  thank Dr. Patrick J. Fox for motivating this work and useful
  discussion on the manuscript.  Finally, we thank the following
  organizations for indispensable logistics and other supports: the
  CJPL administration and the Yalong River Hydropower Development
  Company Ltd.
\end{acknowledgments}

\bibliographystyle{apsrev4-1}
\bibliography{refs.bib}

\end{document}